\documentclass{iau}
\usepackage{graphicx}

\title[IAUS291.~~Pilot pulsar surveys with LOFAR] 
{Pilot pulsar surveys with LOFAR} 

\author[T. Coenen]  
{Thijs Coenen$^1$\\on behalf of the LOFAR Pulsar Working Group}

\affiliation{$^1$Anton Pannekoek Instituut, Universiteit van Amsterdam,\\ 
Postbus 94249, Amsterdam, The Netherlands \\ email: {\tt t.coenen@uva.nl} \\}

\pubyear{2012}
\volume{291}  
\jname{\mbox{Neutron Stars and Pulsars: Challenges and Opportunities after 80 years}}
\editors{J. van Leeuwen, ed.} 
\begin{document}

\maketitle

\begin{abstract}
We are performing two complementary pilot pulsar surveys as part of LOFAR 
commissioning. The LOFAR Pilot Pulsar Survey (LPPS) is a shallow all-sky 
survey using an incoherent combination of LOFAR stations. The LOFAR 
Tied-Array Survey (LOTAS) is a deeper pilot survey using 19 simultaneous 
tied-array beams. These will inform a forthcoming deep survey of the entire 
northern hemisphere, which is expected to discover hundreds of pulsars. Here
we present early results from LPPS and LOTAS, among which are two 
independent pulsar discoveries.
\end{abstract}


\firstsection 
\section{Introduction}\label{section-introduction}
The Low Frequency Array (LOFAR) is a radio telescope under construction in 
Western Europe \cite[(van Haarlem \etal\ in prep.)]{VANHAARLEM}. LOFAR's core is in the province of 
Drenthe, The Netherlands. It operates in two bands, a low band 
($10 - 90$ MHz) and a high band ($110 - 250$ MHz), each with its own 
type of antenna. These bands cover the lowest 4 octaves of the radio window. 
LOFAR is a phased array consisting of many relatively inexpensive antennas, 
whose signals are digitized and then processed mostly in software. The 
antennas are grouped in stations where each station is the equivalent of a 
steerable dish in a traditional interferometer.  LOFAR will consist of 40 
Dutch stations and 8 international stations (in France, Germany, Sweden and 
the United Kingdom). All of the core and international stations are complete, 
and more than half of the remote Dutch stations are also operational. The 
same software correlator allows LOFAR to operate in both imaging and 
beam-formed modes. Each beam in the beam-formed mode is synthesized by adding 
data from multiple stations and acts as the field-of-view of a single-dish 
telescope.  This mode provides the high time resolution required for pulsar
survey observations \cite[(Stappers \etal\ 2011)]{2011AA...530A..80S}. The number of beams that 
LOFAR can create simultaneously is only constrained by compute power and 
system throughput, but can be as high as several hundreds, covering up to 
hundreds of square degrees. In the low band these beams can be pointed 
anywhere on the sky in the high band they need to be clustered.

Low-frequency observations of pulsars are complicated by 3 effects: 
dispersion, scattering and higher sky background temperature. Dispersion, 
the delay caused by free electrons in the interstellar medium (ISM), smears 
out pulsar signals thus making them less detectable. Since the delay scales 
with frequency $\nu$ as $\nu^{-2}$, LOFAR observations are strongly affected. 
Fortunately, the availability of abundant compute power allows incoherent 
de-dispersion with small channel bandwidths, or even online \emph{coherent} 
de-dispersion for up to 40 different trial dispersion 
measures\footnote{Intermediate dispersion measure trials can be filled in 
with offline incoherent de-dispersion.}. Scattering caused by multipath 
propagation in the clumpy ISM is highly dependent on the line-of-sight, and
scales as $\nu^{-4.4}$. It cannot be easily corrected for. This is a problem
especially for pulsar observations in the Galactic plane. Conversely, 
because LOFAR is more strongly affected by scattering, it is an instrument
well suited to studying the structure of the ISM with pulsars. Finally, the
background sky temperature increases with decreasing observing frequency, as
$\nu^{-2.6}$. This effect is mostly a problem towards the Galactic plane. 
Fortunately, two properties of pulsars potentially increase the chance of a 
LOFAR detection. Their spectrum generally goes as  $\nu^{-1.8}$ and their 
beams are broader at low frequencies.

LOFAR is an efficient pulsar surveying instrument. LOFAR's antenna elements
are sensitive to a large part of the sky, and the correlator can create 
many beams with a combined field-of-view of tens to hundreds of degrees. 
Therefore, LOFAR can cover a large part of the sky in little observing time,
and/or use long dwell times. The beam-forming can happen in two modes: 
station data can be combined either coherently or incoherently. The coherent
mode offers maximum raw sensitivity, while the incoherent mode trades 
sensitivity for larger field-of-view; see 
\cite[van Leeuwen \& Stappers (2010)]{2010AA...509A...7V} for the details of this
trade-off. During LOFAR commissioning the LOFAR Pulsar Working Group 
performed pilot surveys in each of these modes. The first such survey, the 
LOFAR Pilot Pulsar Survey (LPPS), used incoherent addition of 7 beams 
created at station level. The second survey, the LOFAR Tied Array Survey 
(LOTAS), exercised the ability to create 19 tied-array beams by coherently 
adding station data. 

In Section \ref{section-lpps} we give an overview of the LPPS survey and 
present some recent results. In Section \ref{section-lotas} we do the same
for the LOTAS survey and finally in Section \ref{section-discussion} we 
discuss the lessons learned and the future outlook for pulsar surveys
with LOFAR.

\section{The LOFAR Pilot Pulsar Survey}\label{section-lpps}
\begin{figure}
\begin{center}
\includegraphics[width=0.75\textwidth]{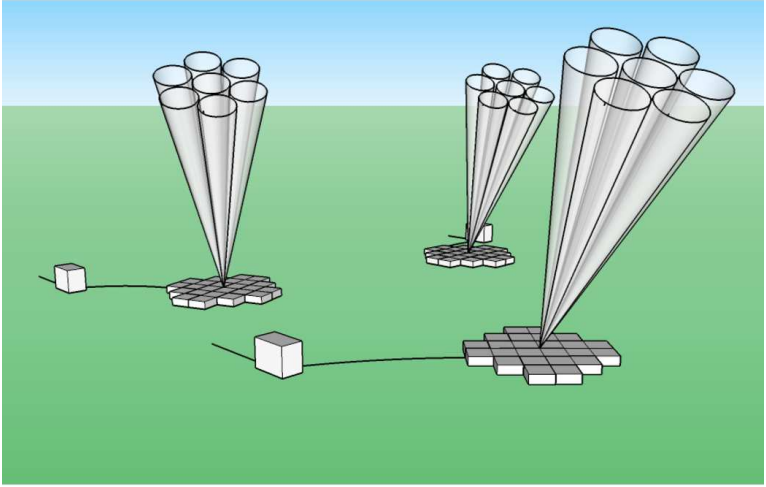}
\caption{Three LOFAR high-band sub-stations in LPPS mode, where they each 
create 7 beams of 6.8 MHz bandwidth each. Since each station is pointed in 
the same direction, LOFAR's central correlator can incoherently combine the
station beams. This maintains the large station beam field-of-view, about 75
square degrees for LPPS, whilst increasing the overall sensitivity with
square root of the number of substations added. The LPPS survey used up to
44 substations per observation.}
\label{figure-lpps-beamforming}
\end{center}
\end{figure}

The LPPS survey was started almost as soon as LOFAR gained the ability to 
form several beams at station level and combine those beams for all stations
at the central correlator. The observations were taken in December 2010 and
early January 2011. Each LPPS pointing had 7 beams with 6.8 MHz bandwidth, a
sampling time of 0.65 ms and a dwell time of 57 minutes. This long dwell time
was possible because each pointing covered about 75 square degrees (see
Figure \ref{figure-lpps-beamforming}). LPPS is comprised of about 250 such 
observations. The search processing is now complete, and candidate 
inspection is in progress.

\begin{figure}
\begin{center}
\includegraphics[width=\textwidth]{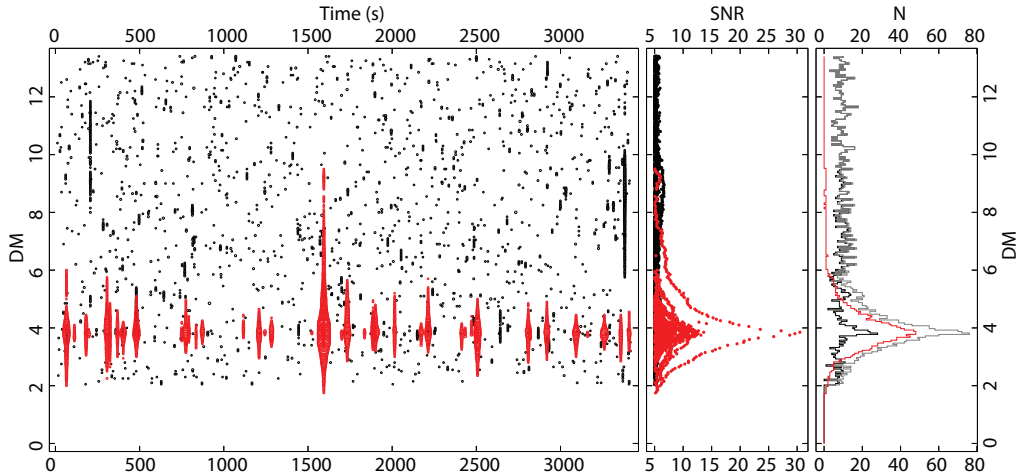}
\caption{
Pulsar PSR~J0240+62, a recent low dispersion measure discovery, was re-detected
in the LPPS survey. The pulsar, as found by our single pulse search 
post-processing algorithms, is shown in red and the noise background
in black.
}
\label{figure-sps-j0240+62}
\end{center}
\end{figure}

The search was performed with a custom Python pipeline using tools from the
PRESTO \cite[(Ransom 2001)]{2001PhDT.......123R} pulsar data reduction package. The data 
were reduced at ASTRON and the University of Manchester. We searched the
data for both periodicities and single dispersed pulses. In Figure 
\ref{figure-sps-j0240+62} we present an interesting single-pulse detection 
of PSR~J0240+62, a recently discovered pulsar with a low dispersion measure (DM)
of $\sim 4$ \cite[(Hessels \etal\ 2008)]{2008AIPC..983..613H}. This shows that LOFAR has the
outstanding ability to detect low-DM sources. The periodicity search yielded
the first independent\footnote{Only weeks before, the GBNCC pulsar survey 
\nocite{GBNCCWEBPAGE} had discovered this pulsar (\emph{priv. comm.}).} 
discovery of a pulsar, PSR~J2317+68, with LOFAR (see the left panel of Figure 
\ref{lpps-lotas-discoveries}).

\section{The LOFAR Tied Array Survey}\label{section-lotas}
LOTAS, the second pulsar commissioning survey, was set up to test the ability
to create multiple tied-array beams for surveying. For this survey, we used 
the 6 inner-most stations to create 19 beams, each with LOFAR's full 
bandwidth of 48\,MHz. The observations were 17 minutes each with a sampling
time of 1.3\,ms. Since these 6 stations act as one large, single station the
beams cover approximately 3.7 square degrees total, a smaller area than 
those for the LPPS survey. For LOTAS we observed about 200 pointings.

\begin{figure}
\centering
\includegraphics[width=0.75\textwidth]{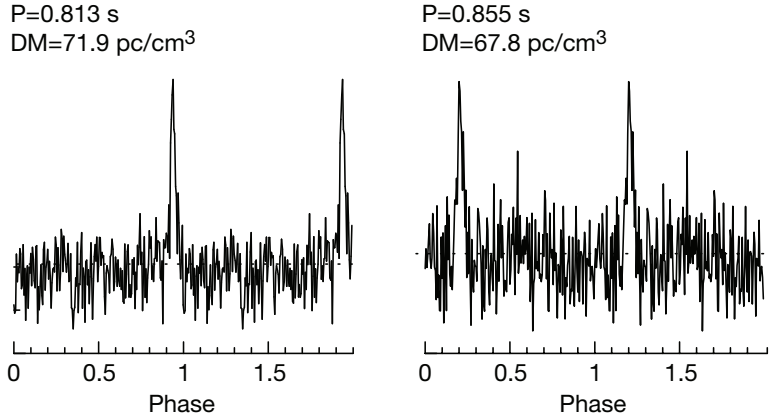}
\caption{The two independent discoveries in the LOFAR pulsar commissioning 
surveys. On the left PSR~J2317+68, found in the LPPS data; on the right
PSR~J2243+69, found in LOTAS.}
\label{lpps-lotas-discoveries}
\end{figure}

The data reduction for this survey is being performed with an updated 
version of the LPPS data reduction pipeline, running at The University of
Manchester and at the SARA Grid Node in Amsterdam. The LOTAS survey has so
far yielded the second, again independent\footnote{This pulsar has also been
discovered recently by the GBNCC survey (\emph{priv. comm.}).}, discovery of
a pulsar, PSR~J2243+69 (see the right panel of Figure 
\ref{lpps-lotas-discoveries}).

\section{Discussion and further work}\label{section-discussion}
The independent discovery of two pulsars in LOFAR's pulsar commissioning 
surveys shows that LOFAR is already competitive for such work. It needs to be
emphasized that the LPPS survey used LOFAR in a very early stage when it was
not yet fully calibrated. The data processing for the LPPS and LOTAS surveys
is not yet complete and we expect that these surveys will yield more 
discoveries soon. For LPPS we are wrapping up the processing and inspecting
the results. Because of the large field-of-view $\times$ time on sky we aim 
to derive a limit on the rate of bright radio bursts, in the absence of 
strong scattering constraints. 

LOFAR's capabilities are still being extended. For targeted observations
its sensitivity will increase because the central single clock will be
rolled out beyond the inner-most 6 stations, connecting a total of 24 core 
stations by the end of 2012 --- providing a 4-fold increase in raw 
sensitivity. Better usage of the available network bandwidth is expected to
increase the observing bandwidth to 80 MHz. Monitoring of the individual 
stations' data quality and a better understanding of the interference 
environment will further increase the data quality compared to that of our
early surveys. 

Now that LOFAR is emerging from its commissioning period, the LOFAR Pulsar
Working Group is gearing up to perform a deeper survey of the northern 
celestial hemisphere. This survey, the LOFAR Tied Array All-Sky survey 
(LOTAAS), will use 61 tied-array beams formed with only the 6 inner-most 
stations. A move to using all core stations would increase raw-sensitivity 
by a factor 4 over this setup. The decrease in field-of-view, however, 
cannot be compensated for by creating more beams (as the computing and 
storage requirements exceed what is available currently or in the near 
future). With a sampling time of 164\,$\mu s$ LOTAAS, unlike LPPS and 
LOTAS, is designed to be sensitive to both regular and millisecond 
pulsars.

\end{document}